# Gravitational Waves II: Emitting Systems.

M. Cattani
Instituto de Fisica, Universidade de S. Paulo, C.P. 66318, CEP 05315-970
S. Paulo, S.P. Brazil .
E−mail: mcattani@if.usp.br

Abstract.
We use the basic equations that predict the emission of gravitational waves according to the Einstein gravitation theory to calculate the luminosities and the amplitudes of the waves generated by binary stars, pulsations of neutron stars, wobbling of deformed neutron stars, oscillating quadrupoles, rotating bars and collapsing and bouncing cores of supernovas. This paper was written to graduate and postgraduate students of Physics.

Resumo.
Usamos as equações básicas que prevêem a emissão de ondas gravitacionais de acordo com a teoria de gravitação de Einstein para calcular as luminosidades e amplitudes de ondas geradas por estrelas binárias, pulsações de estrelas de nêutrons, precessão de estrelas de nêutrons deformadas, quadrupolos oscilantes, barras girantes e nos processos cataclísmicos que dão origem a supernovas. Este artigo foi escrito para os estudantes de graduação e pós−graduação em Física.

## I. Introdução

Todas as teorias relativísticas aceitas sobre gravitação [1-4] prevêem a existência de ondas gravitacionais (OG). No artigo anterior,[5] assumindo a Teoria de Gravitação de Einstein como sendo a mais fidedigna, deduzimos as equações gerais que prevêem a emissão de OG. Essas ondas podem ser interpretadas como enrugamentos na curvatura do espaço tempo 4−dim de Riemann que se propagam com a velocidade da luz. Essas ondas[5] são caracterizadas por duas amplitudes adimensionais $h_+$ e $h_x$ que podem ser entendidas[1-4] como campos escalares no espaço−tempo. Conforme foi visto,[5] o momento de quadrupolo de massa é o primeiro termo na expansão multipolar do campo de radiação.[1-4] O termo quadrupolar será considerado como sendo o dominante nos cálculos que faremos nesse artigo. É muito importante lembrar[1-5] que as equações foram obtidas assumindo que as massas que geram as ondas se movimentam com velocidades muito pequenas comparadas com a velocidade da luz (slow motion) e que os efeitos de auto−gravitação são pequenos (weak self−gravity).

Como veremos a logo abaixo e pode ser visto em vários livros[1–3,6,7] as OG que poderíamos gerar em nossos laboratórios são tão enormemente pequenas que seria impossível com as técnicas atuais detectá–las numa experiência semelhante a efetuada por Hertz no eletromagnetismo. Assim, os astrofísicos têm feito um enorme esforço para estimar as ondas que poderiam ser produzidas em fenômenos astrofísicos.[1–3] A seguir veremos como usar as equações que deduzimos no artigo anterior[5] para calcular a emissão de ondas por vários sistemas emissores. Do mesmo modo que no artigo anterior,[5] procuraremos analisar a emissão de ondas de uma maneira didática e mais simples possível, sem perder o rigor matemático, citando um número mínimo necessário de livros e trabalhos publicados sobre o assunto. Com esse procedimento buscamos tornar o nosso trabalho acessível a alunos de graduação e pós–graduação em Física. Num próximo artigo *Ondas Gravitacionais III: Sistemas de Detecção* analisaremos os sistemas que são usados em nossos laboratórios para detectar as OG.

Como as equações que usaremos nesse artigo são essencialmente as que envolvem o fluxo de energia gravitacional, potência gravitacional irradiada, sua distribuição angular, seus valores médios, temporal e espacial, vamos apresentar algumas grandezas básicas que serão utilizadas nas Seções seguintes levando em conta equações deduzidas antes[5]. Assim, de acordo com as equações[5] (4.11) e (4.12) o fluxo de energia $\Phi_n$ emitida ao longo de um versor genérico **n**, em unidades de (energia/s área), $\Phi_n = F = (d^2E/dt\, dA)$, onde $dA = r^2 d\Omega$, é dado por

$$F = (G/36\pi\, r^2 c^5)\{(1/2)\dddot{Q}_{\alpha\beta}^2 - \dddot{Q}_{\alpha\beta}\dddot{Q}_{\alpha\gamma}n_\beta n_\gamma + (1/4)(\dddot{Q}_{\alpha\beta}n_\alpha n_\beta)^2\} \quad (I.1),$$

onde $Q_{ij}$ é o tensor momento de quadrupolo do sistema emissor[4,5] definido por

$$Q_{ij} = \int \rho_o (3x^i x^j - \delta_{ij} r^2) dV$$

e os pontos sobre Q indicam as derivadas de Q em relação ao tempo.

Como vimos,[4] a energia irradiada em todas as direções, ou seja, a *energia perdida* pelo sistema emissor por unidade de tempo $(-dE/dt)$ é dada por $-dE/dt = (G/45c^5)\dddot{Q}_{\alpha\beta}^2$. A potência total $L_{GW}$ (energia/s) ou *luminosidade gravitacional* do sistema emissor que chega sobre uma superfície esférica que está uma distância r da fonte emissora é dada por

$$L_{GW} = <dE/dt> = r^2 \int <F> d\Omega = (G/45c^5)<\dddot{Q}_{\alpha\beta}^2> \quad (I.2),$$

onde os parênteses $<M>$ indicam uma média temporal da grandeza M.

A energia gravitacional é conduzida, na zona de radiação, por uma onda plana com amplitude $h_o$ e freqüência $\omega$ que estão relacionadas ao valor médio do fluxo $<F>$ através da seguinte equação[4]

$$< F > = (c^3/32\pi G)\, h_o^2\, \omega^2 \qquad (I.3).$$

Levando em conta (I.2) e (I.3) teremos,

$$L_{GW} = < dE/dt > = 4\pi r^2 (c^3/32\pi G)\, h_o^2\, \omega^2 = (G/45c^5) < \dddot{Q}_{\alpha\beta}^{\,2} > \qquad (I.4).$$

Assim, a (I.4) permite−nos estabelecer o seguinte protocolo:
(1) calculamos a luminosidade através da relação (I.2),

$$L_{GW} = (G/45c^5) < \dddot{Q}_{\alpha\beta}^{\,2} > .$$

(2) A partir da luminosidade $L_{GW}$ calculamos a amplitude $h_o$ a uma distância r do centro emissor através de (I.4):

$$h_o^2 = (32\pi G/\omega^2 c^3)\,(L_{GW}/4\pi r^2) \qquad (I.5)$$

Se uma energia total irradiada $Mc^2$ for gerada em um intervalo de tempo muito curto (num pulso) $\Delta\tau$ teremos [6] numa superfície esférica de raio r

$$\int L_{GW}\, dt = Mc^2 = 4\pi r^2 < F > \Delta\tau = (c^3/8G)\, r^2\, h_o^2\, \omega^2 \Delta\tau \qquad (I.6).$$

De (I.6) podemos calcular[6] a amplitude $h_o$ a uma distância r do centro de emissão em termos das seguintes unidades, [r] = 10 kpc , [$Mc^2$] = $10^{-3}\, M_{sol}c^2$ , [$\tau$] = ms e freqüência [f] = kHz:

$$h_o = 10^{-18}\,(1\text{kHz}/f)(10\text{ kpc}/r)(M/10^{-3}\, M_{sol})(1\text{ ms}/\Delta\tau)^{1/2} \qquad (I.7),$$

lembrando que 1 parsec = pc ≈ 3.26 ly . Como ly = anos−luz = 9.46 $10^{15}$ m teremos pc ≈ 30.84 $10^{15}$ m ~3.1 $10^{18}$ cm.

Usando (I.2)−(I.7) vamos estimar e comparar as luminosidades $L_{GW}$ e as amplitudes $h_o$ devido uma fonte astrofísica e devido a uma fonte que poderia ser construídas num em laboratório. Assim, numa primeira aproximação assumiremos[7] que o momento de quadrupolo de um corpo de massa M e raio R seja dado por Q ~ $MR^2$. Se este corpo rígido vibra com simetria não−radial com freqüência $\omega$ podemos escrever[5]

$$\dddot{Q}_{\alpha\beta} \sim MR^2\omega^3 \sim Mv^3/R \qquad (I.8),$$

onde v é a velocidade do material dentro do corpo. Desse modo $L_{GW}$ será dada por [4]

$$L_{GW} \sim GM^2 v^6/R^2 c^5 \qquad (I.9).$$

Com (I.5) e (I.9) obtemos a amplitude da onda a uma distância r da fonte

$$h_o = \{(32\pi G/\omega^2 c^3)(L_{GW}/4\pi r^2)\}^{1/2} \sim (2\sqrt{2}GM/c^2)(v/c)^2/r \qquad (I.10).$$

Se a esfera rígida for, por exemplo, uma estrela de nêutrons com $M \approx 1.4\, M_{sol} \approx 2.8\, 10^{33}$ g, R = 10 km e v/c ~ 0.03 a sua luminosidade, de acordo com (I.9), será $L_{GW} \sim 3\, 10^{48}$ W. Se ela estiver em nossa Galáxia a uma distância r = 10 kpc da Terra a amplitude da onda na Terra, usando (I.10), será $h_o \sim 10^{-20}$. Por outro lado, se a esfera estiver em um laboratório e for feita de um metal muito duro com $M = 4\, 10^3$ kg, R = 2 m e as vibrações se propagarem no material com uma velocidade $v \approx 5\, 10^3$ m/s teremos $L_{GW} \sim 10^{-17}$ W, que é um valor extremamente pequeno. Usando (I.10), a uma distância de r = 1 m da esfera vibrante, a onda gravitacional teria uma amplitude $h_o \sim 10^{-33}$.

Com as técnicas atuais,[1-3,6,7] de interferometria e sólidos ressonantes, as amplitudes mínimas $h_o$(min) capazes de serem detectadas em pulsos de curta duração são da ordem de $h_o$(min) ~ $10^{-21}$–$10^{-22}$. No caso de emissões periódicas $h_o$(min) ~ $10^{-25}$–$10^{-26}$ se for possível fazer uma integração temporal com duração de $10^7$ s.[8] Assim notamos que seria impossível atualmente detectar OG geradas em laboratórios. Entretanto, estima-se que haja um número enorme de sistemas astrofísicos[1-3,6,7] que seriam possíveis fontes de OG com amplitudes detectáveis. Analisaremos somente alguns deles. Assim, na Seção 1 estudaremos a emissão de ondas pelos seguintes sistemas periódicos: *Estrelas Binárias*, *Pulsações em Estrelas de Nêutrons*, *Precessão de Estrelas de Nêutrons Deformadas*, *Quadrupolo Oscilante* e *Barra em Rotação*. Na Seção 2 estudaremos a emissão de OG geradas num pulso em *Supernovas do tipo II*.

Num próximo artigo Ondas Gravitacionais III: Sistemas de Detecção faremos um estudo sobre os equipamentos que são usados e que estão sendo projetados para detectar as OG em nossos laboratórios.

## 1. Sistemas Emissores Periódicos.

(A) *Estrelas Binárias*.

Consideremos duas estrelas compactas de massas $m_1$ e $m_2$ que se movimentam com velocidade angular ω em órbitas circulares em torno de seu centro de massa (CM) comum O (vide Fig.1). As distâncias delas até O, $R_1$ e $R_2$, respectivamente, são constantes.

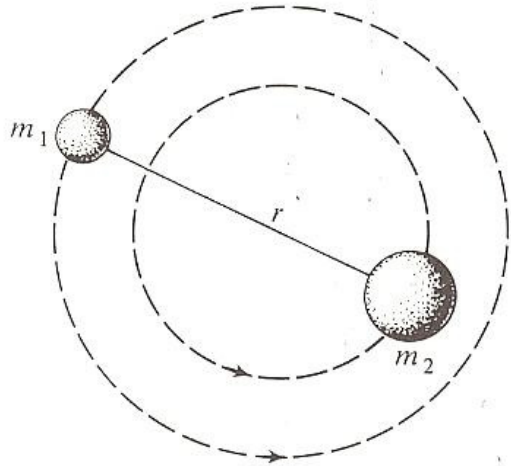

**Figura 1**. Estrelas binárias: duas massas em órbitas circulares em torno do centro de massa delas.[6]

Definindo $r = R_1 - R_2$ a distância relativa entre $m_1$ e $m_2$ e levando em conta que massa reduzida do sistema $\mu = m_1 m_2/M$, onde $M = m_1 + m_2$, vemos que $R_1 = (m_2/M)r$ e $R = -(m_1/M)r$. As coordenadas de 1 e 2 ao longo dos eixos x e y são dadas por $x_1 = R_1 \cos\omega t$, $y_1 = R_1 \sin\omega t$, $x_2 = R_2 \cos\omega t$ e $y_2 = R_2 \sin\omega t$, respectivamente.

Usando a definição de $Q_{ij}$ mostrada em (I.1) e supondo que as massas estão concentradas nos pontos 1 e 2 temos

$Q_{xx} = Q_{xx}^{(1)} + Q_{xx}^{(2)} = 3(m_1 x_1^2 + m_2 x_2^2) - m_1(x_1^2 + y_1^2) - m_2(x_2^2 + y_2^2)$ ou

$Q_{xx} = 3(m_1 R_1^2 + m_2 R_2^2) \cos^2\omega t - (m_1 R_1^2 + m_2 R_2^2)$, ou seja,

$Q_{xx} = 3(m_1 m_2/M) r^2 \cos^2\omega t - (m_1 m_2/M) r^2$.

De modo semelhante obtemos

$Q_{yy} = 3(m_1 m_2/M) r^2 \sin^2\omega t - (m_1 m_2/M) r^2$ e $Q_{zz} = -(m_1 m_2/M) r^2$.

Levando em conta que $\cos^2\omega t = (1 + \cos 2\omega t)/2$, $\sin^2\omega t = (1 - \cos 2\omega t)/2$ e que o termo $(m_1 m_2/M) r^2$ sendo constante não irá contribuir para a emergia emitida consideraremos somente $Q_{xx}$ e $Q_{yy}$:

$Q_{xx} = 3(m_1 m_2/M) r^2 (1 + \cos 2\omega t)/2$

e

$Q_{yy} = 3(m_1 m_2/M) r^2 (1 - \cos 2\omega t)/2$,

que vão dar

$$\ddot{Q}_{xx} = -\ddot{Q}_{yy} = 3\omega^3 2^3 (m_1 m_2/M)\, r^2 \sin 2\omega t \quad e \quad \ddot{Q}_{zz} = 0 \qquad (A.1).$$

Note-se que a dependência temporal dos quadrupolos responsáveis pela emissão da onda, segundo (A.2), é sin2ωt. Isto implica que a freqüência angular $\omega_g$ da onda emitida é $\omega_g = 2\omega$, ou seja, é duas vezes maior do que a freqüência orbital ω.

Levando em conta (A.1) vemos que o valor médio da energia emitida ou *luminosidade gravitacional* $L_{GW} = <dE/dt>$, usando (I.2), será dada por

$$<dE/dt> = -(G/45c^5)\, <\ddot{Q}_{ij}^2> = -(G/45c^5)\, \{<\ddot{Q}_{xx}^2> + <\ddot{Q}_{yy}^2>\},$$

ou seja,

$$L_{GW} = <dE/dt> = -(32G/9c^5)(m_1 m_2/M)^2\, r^4 \omega^6 \qquad (A.2).$$

Uma quantidade apreciável de energia é emitida por estrelas binárias com massas m ~$M_{sol}$ se elas estiverem muito próximas.

Consideremos, por exemplo, o binário W Ursae Majoris (WUMa)[6,7] no qual $m_1 = 0.76\, M_{sol}$, $m_2 = 0.57\, M_{sol}$, $r = 1.5\, 10^{11}$ cm e um período orbital $T = 2\pi/\omega = 8.0$ h. A freqüência $f_g$ da onda emitida é $f_g = 2/T = 1/4 = 0.25$/hr. Usando (1.4) temos $L_{GW} = <dE/dt> \approx 10^{29}$ erg/s $= 10^{22}$ W, ou seja, ~ $10^{-5}$ menor do que a potência luminosa, ou *luminosidade*, do sol. Como a distância de WUMa a Terra é ~ 360 ly ~ 110 pc o fluxo de energia sobre a Terra $\Phi = L_{GW}/4\pi r^2$ é extremamente pequeno, ou seja, $\Phi \sim 10^{-13}$ erg/cm$^2$. A amplitude $h_o$ na Terra da OG proveniente de WUMa, de acordo com a (I.5), seria $h_o \sim 10^{-28}$. Com as técnicas atuais essas ondas não poderiam ser detectadas, pois $h_o(lim) \sim 10^{-22} - 10^{-21}$.

A energia mecânica total *E* do binário é dada, usando o *teorema do virial*,[10] por $E = -(1/2)Gm_1 m_2/r = -(1/2)G\mu M/r$. A energia gravitacional emitida diminui *E* e, consequentemente, leva a um decréscimo da distância r com o tempo. Assim, como r = r(t) e fazendo $dE/dt = L_{GW} = <dE/dt>$ verifica-se que a taxa de redução de r é dada por

$$dr/dt = -(64\mu M^2 G^3/5c^5)/r^3 \qquad (A.3).$$

Para o binário WUMa de acordo com (A.3) temos $dr/dt \approx -10$ cm/y.

Integrando (A.3) obtemos a evolução do raio orbital $r(t) = r_o(1 - t/\tau)$, onde $r_o$ é o *raio atual* e $\tau = (1/4)(-E/L_{GW})$ é o tempo de "espiralamento" ou, simplesmente, *tempo espiral*, isto é, o tempo que a estrelas levam para se encontrarem na origem.

Na pág.990 da livro de Misner, Thorne e Wheeler[3] há uma tabela mostrando cálculos de $L_{GW}$, dr/dt e τ para vários sistemas estelares binários (inclusive Sol & Júpiter), binários de estrelas de nêutrons e de buracos negros. De acordo com essa tabela há vários sistemas binários conhecidos que poderiam ser detectados. Eles têm $L_{GW}$ ~ $10^{30}$ erg/s, T ~ 0.3 day – 81 min e estão a uma distância r ~ 10 – 400 pc da Terra. Nessa mesma tabela são estimados binários hipotéticos de estrela de nêutrons e de buracos negros que poderiam ser detectados e que estão a uma distância r ~ 1 kpc da Terra com $L_{GW}$ ~ $10^{41}$ – $10^{56}$ e T ~ 0.10 ms – 3.2 y. Apenas como referência, a distância do *cluster* de Virgo que contém cerca de 2000 galáxias até a Terra é r ≈ 15 Mpc da Terra. A distância da Terra ao centro de nossa Galáxia é r ≈ 3 $10^4$ ly ≈ 9.78 pc.

*Pulsar PSR 1913+16*

Usaremos agora a (A.2) para explicar o decréscimo observado do decaimento do período orbital T =2π/ω de um sistema binário[7] onde uma das estrelas é o **pulsar PSR 1913+16** (vale a pena lembrar que todos os pulsares conhecidos são estrelas de nêutrons). Ele foi descoberto por Hulse e Taylor[11] usando o radiotelescópio de 305 m em Arecibo. Ambas as estrelas do binário que está localizado a 5 kpc da Terra na constelação de Áquila, têm massas da ordem de 1.4 $M_{sol}$. Para estimarmos dτ/dt suporemos que a energia radiada < dE/dt > = $L_{GW}$ dada por (A.2) (levando em conta que $m_1 = m_2 = m$ e fazendo r = 2a) é proveniente da energia total mecânica E do binário

$$E = mv^2 - Gm^2/2a \qquad (A.4).$$

Usando a equação radial de movimento para cada estrela $mv^2/a = Gm^2/4a^2$ obtemos $v^2 = Gm/4a$ de onde tiramos $\omega^2 = (v/a)^2 = Gm/4a^3$. Substituindo v dada por esta expressão em (A.4) obtemos

$$E = -Gm^2/4a = -(Gm^2/4)(4\omega^2/Gm)^{1/3} \qquad (A.5),$$

da qual deduzimos dE/E = 2dω/ω = − (2/3)(dT/T), onde T = 2π/ω. A quantidade observável (dT/dt)/T, que é a razão de decaimento do período orbital é então dada por

$$(dT/dt)/T = (3/2)(dE/dt)/E = (3/2)(L/E) = -(768/5)(\omega^6 a^5/c^5).$$

Substituindo nesta equação ω que é dada por ω = (v/a) = $(Gm)^{1/2}/2a$ temos finalmente

$$(dT/dt)/T = -(12/5)(G^3m^3/c^5a^4) \qquad (A.6).$$

Como a órbita é elíptica [7, 11–13] com excentricidade *e* = 0.617127 o lado direito de (A.6) precisa ser multiplicado por um fator de correção [13]

α = (1+73e$^2$/24+37e$^4$/96)/(1−e$^2$)$^{7/2}$ ≈ 11.856. Introduzindo esse fator α e usando os parâmetros orbitais que foram medidos [7,11−13] com grande precisão verifica−se que

$$(dT/dt)_{teor} = -2.403(2) \times 10^{-12}$$

Como as medidas efetuadas [7,11−13] dão

$$(dT/dt)_{exp} = -2.40(9) \times 10^{-12}$$

notamos que há um excelente acordo, diríamos impressionante, entre os resultados experimentais e os teóricos. Isso tem um duplo significado.[7] Ele mostra uma forte confirmação da Teoria de Gravitação fora do sistema solar e é uma evidência indireta da existência das OG.

Muitas informações sobre o binário *PSR 1913 + 16* e a comparação entre $(dT/dt)_{teor}$ e $(dT/dt)_{exp}$ em um gráfico em função de t podem ser vistos, por exemplo, no livro do Kenyon[7] e nas referências [11−13] citadas acima.

Calculemos a luminosidade do binário *PSR 1913+16*, dada pela (A.2), $L_{GW} = (32G/9c^5)(m_1 m_2/M)^2 r^4 \omega^6$. Levando em conta que $m_1 = m_2 \approx 1.4 M_{sol} \approx 2.8 \times 10^{33}$ g, o período orbital $T = 2\pi/\omega \approx 7.75$ h $= 2.79 \times 10^4$ s e que a distância entre as estrelas é $r \sim 10^6$ km obtemos $L_{GW} \sim 10^{29}$ W. Assim, usando a (I.5), levando em conta que a distância da Terra ao pulsar é $r \sim 5$ kpc $= 15.5 \times 10^{18}$ cm e que a ω da onda é $4\pi/T$, amplitude $h_o$ da onda que chega à Terra é $h_o \sim 6.4 \times 10^{-21}$.

*(B) Pulsações em Estrelas de Nêutrons.*

Muitos modelos foram propostos [1,2,14−17] para determinar a equação de estado do núcleo das estrelas de nêutrons. Uma força repulsiva gerada por um mar fermiônico de nêutrons equilibra a energia gravitacional da matéria estelar. Numa primeira aproximação podemos descrever essa estrela como sendo um núcleo gigante composto de nêutrons com um raio R ~ 10 km e massa M ~ $M_{sol}$; com uma densidade constante ρ ~ $10^{14} - 10^{15}$ g/cm$^3$, comparável com a densidade da matéria nuclear. Nesse caso a equação de estado é calculada assumindo que os nêutrons constituem um gás ideal[18] degenerado de fermions. Equações de estado mais realísticas são obtidas levando em conta a interação nuclear de curto alcance entre os nêutrons e tratando a matéria neutrônica como sendo um líquido auto−gravitante que, em determinadas condições, poderia se solidificar. Muitos artigos [14−19] mostram que é plausível esperar que a matéria neutrônica auto−gravitante seja governada por uma equação de estado que é muito próxima da de um sólido elástico. Evidências do comportamento viscoelástico são obtidas através da observação de *glitches* em pulsares.[20] Nessas condições, vamos estimar as freqüências de pulsação ω da estrela quando o seu estado de equilíbrio é perturbado levando em conta que a compressibilidade $\chi$ da matéria neutrônica é dada por $\chi = -(\Delta\rho/\Delta p)/\rho$ onde p é a sua pressão e $\rho$ a sua densidade. Lembrando que a velocidade $v_s$ de propagação das ondas sonoras é dada por[21,22] $v_s = 1/\sqrt{\chi\rho} = \sqrt{\partial p/\partial \rho}$ ela pode ser escrita, numa primeira aproximação, como $v_s \sim \sqrt{p/\rho}$. Como a pressão típica p em uma esfera auto-gravitante[23] de raio R e massa M é $p = GM^2/R^4$ obtemos $v_s \sim (G/\rho)^{1/2} (M/R^2) \sim (GM/R)^{1/2}$. Assumindo que os comprimentos de onda $\lambda$ das oscilações sejam dados por $\lambda \sim R$ devemos ter $\omega = (2\pi/\lambda) v_s \sim 2\pi (G\rho)^{1/2}$. Como $\rho \sim 10^{14}$ g/cm$^3$ constatamos que as freqüências $f \sim 10^3$ Hz. Cálculos mais precisos[1] mostram que as f estão num intervalo que vai de $10^{-2}$ a $10^4$ Hz.

Se as pulsações não-radiais que criam uma deformação temporal $\delta R(t)$ do raio R geram um quadrupolo Q(t) ele pode ser aproximado por $Q(t) \sim MR^2(\delta R(t)/R)$. No caso de uma perturbação periódica, $\delta R(t) = \delta R \cos(\omega t)$ teremos ,

$$Q(t) \sim MR^2(\delta R/R) \cos(\omega t) \qquad (B.1)$$

Como $<\ddot{Q}^2> \sim M^2R^4 \omega^6 (\delta R/R)^2$ a luminosidade $L_{GW}$ gerada pelas pulsações, usando (I.2), será dada por

$$L_{GW} \sim (GM^2R^4\omega^6/c^5)(\delta R/R)^2 \qquad (B.2).$$

Assumindo que $M \sim 1.4$ $M_{sol}$, $R \sim 10$ km e $\omega \sim 10^4$ rad/s teremos $L_{GW} \sim 2 \cdot 10^{55}(\delta R/R)^2$ W . Se as amplitudes de pulsações forem da ordem de $\delta R/R \sim 10^{-2}$ teremos uma luminosidade $L_{GW} \sim 2 \cdot 10^{51}$ W.

No caso de *pulsares*, que são estrelas de nêutrons com velocidades de rotação muito altas, a deformação (vide Seção C) da estrela devido à rotação ampliaria o momento de quadrupolo e, consequentemente, a $L_{GW}$.

Se a estrela de nêutrons pulsante estiver em nossa Galáxia, a uma distância $r \sim 10$ kpc da Terra, a amplitude da onda $h_o$, dada pela (I.5), seria $h_o \sim 10^{-19}$. Se ela estiver no cluster de Virgo, a uma distância $r \sim 15$ Mpc de nós, a amplitude $h_o$ seria $h_o \sim 10^{-21}$.

*(C) Precessão de Estrelas de Nêutrons Deformadas.*

Consideremos uma estrela de nêutrons achatada girando em torno de seu eixo de simetria z com uma velocidade angular $\Omega = 2\pi f$, analogamente ao que acontece com a Terra que tem um spin em torno do eixo geográfico Norte–Sul. Devido a esse achatamento a estrela teria um momento de quadrupolo. Suponhamos que por alguma razão o eixo z precessiona em torno de um eixo fixo $\zeta$ formando com ele um ângulo $\theta$ (*wobble angle*). Se $\theta \ll 1$ a *wobbling star* emite OG com freqüência muito próxima de f com uma amplitude $h_o$ dada por[24,25]

$$h_o = 1.4 \; 10^{-18} \; \varepsilon \; \theta \; (I_{zz}/10^{45} \text{ g cm}^2) \; f^2_{kHz} \; r^{-1}_{kpc} \quad \text{(C.1)},$$

onde $\varepsilon = (I_{zz} - I_{xx})/I_{zz}$, $I_{kk}$ são os momentos de inércia da estrela e r é distância da estrela ao observador.

Supondo que $M = 1.4 \; M_{sol}$ e que a matéria estelar obedeça a uma equação de estado do tipo Bethe–Johnson I os momentos de inércia $I_{zz}$, $I_{xx}$ e $\varepsilon$ foram calculados[26] em função de $\Omega$ no intervalo $3000 \leq \Omega \leq 6203$ rad/s. Levando em conta, por exemplo, os valores máximo $\Omega = 6203$ rad/s e mínimo $\Omega = 3200$ rad/s e supondo que a estrela esteja dentro de nossa Galáxia a uma distância r ~ 20 kpc da Terra obtemos, usando a (C.1), as seguintes amplitudes em função do *wobble angle* $\theta$ :

$$(h_o)_{min} \sim 10^{-16} \theta \quad \text{e} \quad (h_o)_{max} \sim 10^{-14} \theta$$

Notamos que mesmo para valores muito pequenos de $\theta$ dentro de num intervalo $10^{-5} \leq \theta \leq 10^{-3}$ as OG geradas pela precessão poderiam ser detectadas.[25]

*(D) Barra em Rotação.*

Consideremos uma barra cilíndrica com comprimento L e área da base A, com massa M, que gira num plano (x,y) com freqüência angular $\omega$ em torno do eixo z que passa pelo seu centro de massa O. Vamos indicar por $\xi$ as distâncias dos elementos de massa $dM = \rho_o dV = \rho_o A d\xi$ ao centro O; $\xi$ varia de $-L/2$ até $L/2$. Como as coordenadas (x,y) da barra são dadas por $x(t) = \xi \cos(\omega t)$ e $y(t) = \xi \sin(\omega t)$ os momentos de quadrupolo $Q_{ij}$ da barra, definidos por (I.1)), são dados por

$$Q_{xx} = \int \rho_o (3x^2 - r^2) dV = \rho_o A \int [2\xi^2 \cos^2(\omega t) - \xi^2 \sin^2(\omega t)] d\xi =$$

$$= 2A\rho_o (L/2)^3 [2\cos^2(\omega t) - \sin^2(\omega t)]/3 = A\rho_o (L/2)^3 [1 + \cos(2\omega t)]/3$$

$$= ML^2 [1 + \cos(2\omega t)]/24.$$

Analogamente temos $Q_{yy} = ML^2 [1 - \cos(2\omega t)]/24$. Como $z^i = 0$ temos, finalmente, $Q_{zz} = -ML^2/24$.

Assim, usando os $Q_{ij}$ vistos acima obtemos $<\dddot{Q}_{ij}^2> = M^2 L^4 \omega^6/9$ e desse modo a luminosidade da barra que gira, usando (I.4), é dada por

$$L_{GW} = (G/45c^5) <\dddot{Q}_{\alpha\beta}^2> = (G/45c^5)(ML^2/3)^2 \omega^6 \quad \text{(D.1)}.$$

Suponhamos que a barra seja de aço[3] com $M = 500$ ton $= 5 \; 10^8$ g, $L = 20$ m e $\omega \sim 20$ rad/s de tal modo que ela não se rompa com a alta rotação. Nessas condições vemos que $L_{GW} \sim 10^{-24}$ W. A uma distância de 1 m da barra, usando (I.5), teríamos $h_o \sim 10^{-41}$. Estes valores tão pequenos mostram que a construção de um gerador desse tipo de OG em laboratórios é um empreendimento nada atrativo. Precisamos inventar um outro método mais eficiente para produzir OG no laboratório. Até lá a única alternativa que temos, tendo em vista que as técnicas atuais só nos permitiriam detectar amplitudes $h_o \geq 10^{-22} - 10^{-21}$, é a de considerar fontes astrofísicas.

*(E) Emissão por Quadrupolo Oscilante.*

Um dos exemplos mais simples de sistemas que podem emitir OG é o quadrupolo linear (vide Fig.2) formado por duas massas iguais ($m_1 = m_2 =$ m) conectadas por uma mola.[6] As posições das massas $m_1$ e $m_2$ em relação à origem O são $z_1 = b + a \sin\omega t$ e $z_2 = -b - a \sin\omega t$, onde assumiremos que $b >> a$, de tal modo que $z^2 \approx b^2 + 2ab \sin\omega t$.

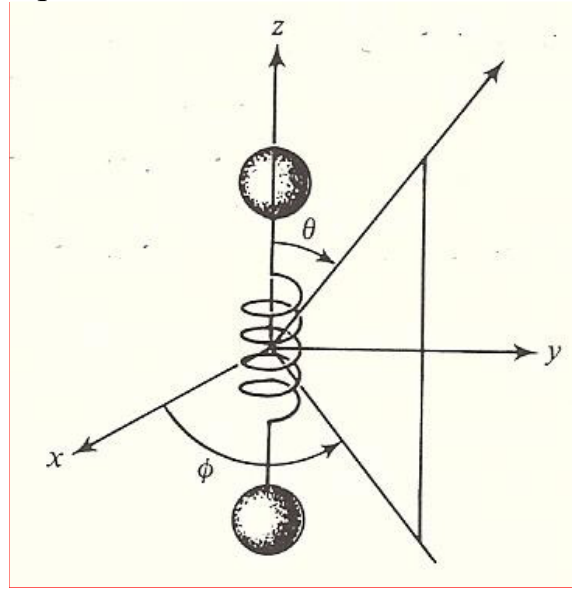

**Figura 2**. Quadrupolo oscilante[6]: duas massas oscilando harmonicamente com amplitude a, em torno dos pontos de equilíbrio $z = \pm b$.

O momento de quadrupolo (vide (I.1)), $Q_{ij} = Q_{ij}\delta_{ij}$ para os tempos retardados é dado por

$Q_{11} = Q_{xx}^{(1)} + Q_{xx}^{(2)} = 0 - 2mb^2[1 + (2a/b) \sin\omega(t-r/c)]$

$Q_{22} = Q_{yy}^{(1)} + Q_{yy}^{(2)} = 0 - 2mb^2[1 + (2a/b) \sin\omega(t-r/c)]$ e     (E.1)

$Q_{33} = Q_{zz}^{(1)} + Q_{zz}^{(2)} = \phantom{0 -} 4mb^2[1 + (2a/b) \sin\omega(t-r/c)]$

ou , simplesmente,

$$Q_{ij} = [1+ (2a/b) \sin\omega(t-r/c)]\, Q_{ij}(0), \qquad (E.2)$$

onde $Q_{11}(0) = Q_{22}(0) = -2mb^2$ e $Q_{33}(0) = 4mb^2$.

Assim, a energia irradiada $d^2E/dtd\Omega$ é dada por (I.1)

$$d^2E/dtd\Omega = (G/36\pi c^5) \left\{ (1/2)\, \dddot{Q}_{\alpha\beta}^{\,2} - \dddot{Q}_{\alpha\beta}\dddot{Q}_{\alpha\gamma}\, n_\beta n_\gamma + (1/4)(\dddot{Q}_{\alpha\beta} n_\alpha n_\beta)^2 \right\} \quad (E.3)$$

onde $n_1 = \sin\theta \cos\varphi$, $n_2 = \sin\theta \sin\varphi$ e $n_3 = \cos\theta$. Assim, usando (E.3) temos

$$d^2E/dtd\Omega = (G/36\pi c^5)\{(1/2)(\dddot{Q}_{11}^{\,2} + \dddot{Q}_{22}^{\,2} + \dddot{Q}_{33}^{\,2}) -$$

$$- (\dddot{Q}_{11} n_1)^2 - (\dddot{Q}_{22} n_2)^2 - (\dddot{Q}_{33} n_3)^2$$

$$+ (1/4)(\dddot{Q}_{11} n_1 n_1 + \dddot{Q}_{22}\, n_2 n_2 + \dddot{Q}_{33}\, n_3 n_3)^2 \} \quad (E.4)$$

Substituindo em (2.4) os versores $n_i$ ($i =1,2,3$) dados em (E.3) e levando em conta que $\dddot{Q}_{ij} = -(2a/b)\,\omega^3 \cos\omega(t-r/c)\, Q_{ij}(0)$ obtemos,

$$d^2E/dtd\Omega = -(G/\pi c^5)[mab\omega^3 \cos(t-r/c)]^2 \sin^4\theta \qquad (E.5)$$

A distribuição angular da radiação gravitacional quadrupolar é mostrada na Fig.3. As ondas são emitidas preferencialmente na direção perpendicular ao eixo z, com simetria cilíndrica em torno de z. A sua distribuição não tem nenhuma semelhança com a quadrupolar elétrica [27,28] lembrando, porém, vagamente a dipolar elétrica [27,28] que tem uma distribuição angular $\sim \sin^2\theta$.

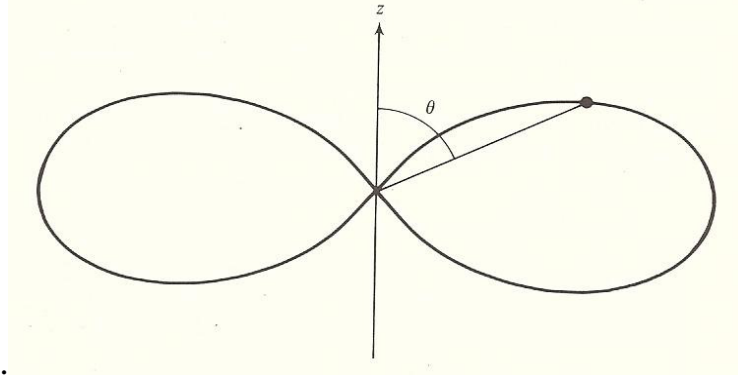

**Figura 3.** Distribuição angular $\sin^4\theta$ das OG geradas pelo quadrupolo oscilante[6] visto na Fig.2.

Embora a (E.5) tenha sido derivada para duas massas pontuais conectadas por uma mola, com ela podemos estimar a ordem de grandeza de OG emitidas pela vibração elástica longitudinal de uma barra.[6]

## 2. Sistemas Emissores de Pulsos de Ondas.

Em muitos processos astrofísicos as OG poderiam ser emitidas[1]em períodos de tempo muito curtos $\Delta\tau$, ou seja, em pulsos, tais como explosões de supernovas, emissão de neutrinos em supernovas, tremores núcleos de estrelas de nêutrons e nascimento de buracos negros. Veremos só o primeiro caso.

*Supernova do tipo II.*

De acordo com a teoria vigente de evolução estelar,[6,7,29] estrelas com massas M ~ 20 ~30 $M_{sol}$ são geradas pela contração gravitacional de gases, principalmente hidrogênio. Devido essa contração a pressão interna e a temperatura podem crescer a tal ponto que dão origem à reações de fusão termonucleares. A energia liberada nesse processo, aumentando a pressão e e a temperatura da matéria estelar, impedem a estrela de colapsar. O hidrogênio se converte em hélio, mas esgotado o hidrogênio a energia produzida diminui e a estrela começa a contrair. Isto faz com que a pressão e a temperatura comecem de novo a crescer dando início a reações termonucleares que convertem o hélio em carbono. Estágios subseqüentes de queima de combustível nuclear produzem, finalmente depois de ~$10^9$ anos, uma estrela com um caroço formado por núcleos de ferro.Como os núcleos de ferro têm a maior energia de ligação por núcleo nem fusões ou fissões podem produzir energias adicionais. As camadas externas da estrela por não terem atingido o estágio do ferro continuam produzindo energia por fusões. O caroço é formado por núcleos de ferro e de elétrons arrancados de núcleos que se movem quase livremente através do volume da estrela. A pressão de ponto–zero do gás degenerado de elétrons[23] é responsável pela maior parte da pressão do caroço e os núcleos dão a maior contribuição para a sua massa. As equações de estado (pressão em função da densidade) que são desenvolvidas baseadas nesse modelo mostram que pode haver equilíbrio estável da estrela se a sua massa não ultrapassar um valor crítico denominado de massa limite M(Chandrasekar) = 1.44 $M_{sol}$.

Assim, se M < 1.44 $M_{sol}$ a estrela fica num estado de equilíbrio, ou seja, a contração cessa e ela se torna uma *Anã Branca*. Uma anã branca típica tem um raio R ~ raio da Terra e densidade ~$10^2$ até $10^4$ kg/$m^3$. A luminosidade é gerada nas camadas superiores da estrela e à medida que vai acabando o combustível nuclear ela é devida à energia térmica residual. À medida que a energia térmica vai sendo perdida por radiação ela vai ficando cada vez menos luminosa transformando–se, após um período de ~$10^{10}$ y, numa estrela apagada ou *Anã Negra*. Devido à instabilidades, à medida que se esfriam, podem se tornar *anãs–brancas pulsantes*.

Se M > 1.44 $M_{sol}$ não há equilíbrio estável, a estrela continua a se contrair, os elétrons no caroço são absorvidos pelos prótons dos núcleos de ferro transformando–os em nêutrons. A neutronização é rápida, levando cerca de 1 s, acompanhada de um colapso do caroço e de uma violenta emissão de neutrinos.[30] Este fenômeno é conhecida como *Processo Urca*. Ele foi discutido pela primeira vez por G.Gamow e M.Schemberg enquanto visitavam o casino da Urca no Rio de Janeiro. Relata−se que Gamow teria dito a Schemberg : a energia desaparece no núcleo de uma supernova tão rapidamente quanto o dinheiro do jogador desaparece na mesa da roleta. O caroço fica muito menor, a sua densidade aumenta muito, tornando−se muito duro devido à repulsão nuclear. As partes externas da estrela implodem e ao se chocarem com o caroço de nêutrons são ricocheteadas com tal violência que uma parte da massa da estrela é ejetada dando origem ao que chamamos de explosão de *supernova tipo II*. Uma supernova observada recentemente é a SN1987a. Estima−se que na explosão a energia liberada esteja entre $10^{-5}$ $Mc^2$ e $10^{-2}$ $Mc^2$. Após a explosão, se a massa residual $M_r$ da estrela[6,7,29] for $M_r < 2$ a $3$ $M_{sol}$ ela pára de se contrair tornando−se uma *Estrela de Nêutrons* com raio R~10 km e densidade $\rho \sim 10^{14} - 5\ 10^{15}$ g/cm$^3$. Se $M_r > 3$ $M_{sol}$ a implosão não consegue ser detida e ela se transforma num "buraco negro".

Assumindo que a massa original de gases tivesse um *spin* no final do processo de enorme contração da estrela a sua velocidade angular seria extremamente alta. Isso acentuaria a assimetria da estrela[25,26] favorecendo a emissão de OG no colapso. Como a astrofísica desses processos é muito complicada só é possível estimarmos alguns limites dos parâmetros[1,6,7] relevantes para radiação gravitacional gerada no colapso. Assim, uma estimativa é obtida assumindo que as OG seriam emitidas num pulso com duração de tempo $\Delta\tau \sim 10^{-3}$ s, carregariam uma energia $\sim 10^{-3}$ $M_{sol}$ e teriam freqüências f $\sim 1/\Delta\tau$ no intervalo de 10 a $10^4$ Hz.

Nessas condições podemos calcular a amplitude $h_o$ da onda que seria detectada na Terra a uma distância r da supernova usando a (I.7)

$$h_o = 10^{-18}\ (1kHz/f)(10\ kpc/r)\ (M/10^{-3}\ M_{sol})(1\ ms/\Delta\tau)^{1/2}\quad .$$

Supondo que f = 1khZ, M = $10^{-3}$ $M_{sol}$, $\Delta\tau$ = 1ms e que r ~10 kpc, ou seja, que a supernova esteja dentro de nossa Galáxia obtemos $h_o$(Galáxia) ~ $10^{-18}$. Explosões de supernovas do tipo II ocorrem uma vez a cada 30 anos em nossa Galáxia. Se a supernova estiver no cluster da Virgem (*Virgo*) ou seja, r ~15 Mpc teríamos $h_o$(Virgo) ~ 7 $10^{-22}$ , ainda possível de ser detectada. Como Virgo contém cerca de 2000 galáxias teríamos uma taxa de uma explosão a cada poucos dias o que aumentaria muito as chances de detecção das OG.